\documentclass[%
 amsmath,amssymb,
 aps,
 prd,
twocolumn, superscriptaddress
]{revtex4-1}

\usepackage[normalem]{ulem}
\usepackage{cancel}

\usepackage[dvipsnames]{xcolor}
\usepackage{graphicx}
\usepackage{dcolumn}
\usepackage{bm}
\usepackage{xcolor}
\usepackage{orcidlink}
\hypersetup{colorlinks, citecolor=green}
\hypersetup{colorlinks=true, linkcolor=Bittersweet, urlcolor=MidnightBlue}
\usepackage{hyperref}
\usepackage{footnote}

\begin{document}

\title{Neutron stars in $f(R,L_m,T)$ gravity}

\author{Cl{\'e}sio E. Mota}
 \email{clesio200915@hotmail.com}
 \affiliation{Departamento de F\'isica, CFM - Universidade Federal de Santa Catarina; C.P. 476, CEP 88.040-900, Florian\'opolis, SC, Brasil.}

\author{Juan M. Z. Pretel}
 \email{juanzarate@cbpf.br}
 \affiliation{Centro Brasileiro de Pesquisas F{\'i}sicas, Rua Dr.~Xavier Sigaud, 150 URCA, Rio de Janeiro CEP 22290-180, RJ, Brazil
}

\author{César O. V. Flores}
 \email{cesarovfsky@gmail.com}
 \affiliation{Centro de Ci\^encias Exatas, Naturais e Tecnol\'ogicas, CCENT - Universidade Estadual da Regi\~ao Tocantina do Maranh\~ao; C.P. 1300,\\ CEP 65901-480, Imperatriz, MA, Brasil
}
 \affiliation{Departamento de F\'isica, CCET - Universidade Federal do Maranh\~ao, Campus Universit\'ario do Bacanga; CEP 65080-805, S\~ao Lu\'is, MA, Brasil
}

\begin{abstract}
This study explores the behavior of compact stars within the framework of $f(R,L_m,T)$ gravity, focusing on the functional form $f(R,L_m,T) = R + \alpha TL_m$. The modified Tolman-Oppenheimer-Volkoff (TOV) equations are derived and numerically solved for several values of the free parameter $\alpha$ by considering both quark and hadronic matter --- described by realistic equations of state (EoSs). Furthermore, the stellar structure equations are adapted for two different choices of the matter Lagrangian density (namely, $L_m= p$ and $L_m= -\rho$), laying the groundwork for our numerical analysis. As expected, we recover the traditional TOV equations in General Relativity (GR) when $\alpha \rightarrow 0$. Remarkably, we found that the two choices for $L_m$ have appreciably different effects on the mass-radius diagrams. Results showcase the impact of $\alpha$ on compact star properties, while final remarks summarize key findings and discuss implications, including compatibility with observational data from NGC 6397's neutron star. Overall, this research enhances comprehension of $f(R,L_m,T)$ gravity's effects on compact star internal structures, offering insights for future investigations.

\end{abstract}
\maketitle

\section{Introduction}

Einstein's theory of gravity proposed over a century ago not only aids us in understanding various aspects of the universe but also continues to undergo significant experimental testing. These tests include the precession of Mercury's perihelion \cite{1915SPAW.......831E}, accurately predicted, recent detections of gravitational waves originating from binary black hole systems \cite{LIGOScientific:2016aoc}, and neutron star (NS) mergers \cite{LIGOScientific:2017vwq} observed by the Virgo and LIGO collaboration (Laser Interferometer Gravitational-Wave Observatory), along with the first image of a black hole's shadow obtained by the Event Horizon Telescope project \cite{EventHorizonTelescope:2019dse}. Consequently, the results predicted by General Relativity (GR) are in excellent agreement with observational data collected, primarily since the early 20th century.

Despite the success of GR, in recent years, there have been proposals for alternative theories of gravity, often referred in the literature as modified gravity theories \cite{Nojiri:2003ft, Allemandi:2005qs, Nojiri:2017ncd}. Some of these theories seek to extend GR by introducing additional terms into the standard Einstein-Hilbert action. An example of this is $f(R)$ gravity \cite{Buchdahl:1970ynr, DeFelice:2010aj, Capozziello:2011et, Nojiri:2010wj, Sotiriou:2008rp} and its various extensions \cite{Harko:2011kv, Harko:2010mv, Odintsov:2013iba}, where $R$ is the Ricci scalar. Some arguments in favor of these theories suggest that the rotation speed of spiral galaxies, represented by the rotation curve, could be explained without resorting to the presence of dark matter. Additionally, it is proposed that the accelerated expansion of the universe could be understood without the inclusion of dark energy, through the updating of the theory of gravity beyond GR \cite{Olmo:2006zu, Hui:2009kc, Bertolami:2007gv, Harko:2010vs, Abdalla:2004sw, Cognola:2007zu}. Indeed, it is expected that the adopted gravitational model be capable of describing all available astronomical observations, among which NSs and quark stars (QSs) stand out.

The gravitational environments surrounding NSs, exotic stars, and the proximity of black hole event horizons pose unique challenges and offer singular opportunities \cite{Chandrasekhar1931, Lattimer:2004pg, Will:2014kxa}. Shaped by the immense mass and density of compact celestial objects, these extraordinary scenarios establish a unique platform where gravity manifests with unparalleled intensity. These environments evolve into observational arenas, allowing for the direct observation of gravitational forces to conduct theoretical tests of gravity.

Among all the observables of a compact star we have the mass, radius and the gravitational redshift. The mass and radius are very important, because these quantities can be obtained directly from different astronomical observations. In fact, their values offer significant information about the global structure of the star and, at the same time, we can obtain improved insights into the equation of state that governs the microscopic interior of the star. It is important to mention that the recent launched telescopes have more sensitivity and therefore it is possible to obtain more precise values for the mass and radius of a NS and strange quark stars. Moreover, thanks to this new technology it is possible to investigate the validity of GR (as well as its modifications) in the strong-field regime. Nowadays, it has become very common to explore the extreme physics of NSs taking into account the mass-radius window imposed by data from the Neutron Star Interior Composition Explorer (NICER) \cite{Miller2021}, MM-Newton Data and the gravitational wave laser interferometers LIGO-Virgo \cite{LIGOScientific:2017vwq}. Therefore, we can say that we are in a era where it is possible to use advanced observations to obtain more information about compact stars and explore the possible modifications of conventional GR.

In this study we will focus on $f(R,L_m, T)$ theory of gravity \cite{Haghani:2021fpx}, which generalizes and unifies the $f(R,T)$ and $f(R,L_m)$ gravity models. In other words, this theory consistently impose that the gravitational Lagrangian is an arbitrary function of the Ricci scalar, the trace of the energy-momentum tensor and of the matter Lagrangian density. Particularly, we will adopt the $f(R,L_m,T) = R+ \alpha TL_m$ model and investigate the effect of the free parameter $\alpha$ on the most basic macroscopic properties of a compact star such as radius, mass and the surface gravitational redshift. To do so, it becomes necessary to obtain the modified stellar structure equations for the two different choices of $L_m$.

We will specifically examine the effects of $f(R,L_m, T)$ theory on the internal structure of relativistic compact stars made of hadronic matter for NSs and quark matter for QSs. For this purpose in section \ref{sec2}, we will, at first, give a little review of how the gravitational field equations are obtained in $f(R,L_m, T)$ theory of gravity. Then in section \ref{sec3}, from the field equations, we will obtain a modified version of the well-know Tolman-Oppenheimer-Volkoff (TOV) equations for the hydrostatic equilibrium state of a compact star. In section \ref{sec4} we will compute the mass versus radius relation and discuss our outcomes in terms of $\alpha$. Finally, in section \ref{sec5}, we give some remarks and some perspectives.

\section{Gravitational field equations in $f(R,L_m,T)$ theories} \label{sec2}

Proposed by Haghani and Harko \cite{Haghani:2021fpx} as a generalization and unification of the $f(R,T)$ \cite{Harko:2011kv} and $f(R,L_m)$ \cite{Harko:2010mv} gravity models, the gravitational Lagrangian is given by an arbitrary function of the Ricci scalar $R$, of the trace $T$ of the energy-momentum tensor $T_{\mu\nu}$, and of the matter Lagrangian $L_m$, so that $L_{grav}=f(R,L_m,T)$. Thus, the full action in $f(R,L_m,T)$ gravity theories reads
\begin{equation}
    S = \frac{1}{16 \pi} \int f(R,L_m,T) \sqrt{-g} d^4x + \int L_m \sqrt{-g}d^4x,
    \label{action}
\end{equation}
where $g$ is the determinant of the metric tensor $g_{\mu\nu}$, with Greek indices assuming the values $0-3$. Throughout this work we will use geometrized units, that is, $c = G = 1$ and we use the metric signature $(-,+,+,+)$. However, our results will be given in physical units for comparison purposes.

Now, varying the action \ref{action} with respect to the inverse metric $g^{\mu\nu}$, one obtains the following field equations in $f(R,L_m,T)$ gravity
\begin{align}
    f_RR_{\mu\nu} &- \frac{1}{2} [f-(f_L + 2f_T)L_m] g_{\mu\nu} + (g_{\mu\nu}\Box -\nabla_{\mu} \nabla_{\nu})f_R  \nonumber  \\
    &= \left[8\pi + \frac{1}{2}(f_L + 2f_T)\right] T_{\mu\nu} + f_T\tau_{\mu\nu},  \label{eq8}
\end{align}
in which $\Box \equiv \partial_{\mu}(\sqrt{-g}g^{\mu\nu}\partial_{\nu})/\sqrt{-g}$, $f_{R}\equiv \partial f/ \partial R$, $f_{T}\equiv \partial f/ \partial T$, $f_{L} \equiv \partial f/ \partial L_{m}$, $R_{\mu\nu}$ is the Ricci tensor,  $\nabla_{\mu}$ the covariant derivative with respect to the symmetric connection associated to $g_{\mu\nu}$, and the new tensor $\tau_{\mu\nu}$ is defined as \cite{Haghani:2021fpx}
\begin{equation}\label{EqTau}
\tau_{\mu\nu} = 2g^{\alpha\beta} \frac{\partial^{2}L_m}{\partial g^{\mu\nu} \partial g^{\alpha \beta}}.
\end{equation}
It is evident that, when $f(R,L_m,T)=f(R)$, from Eq.~(\ref{eq8}) we retrieve the field equations in the framework of metric $f(R)$ gravity \cite{Nojiri:2009kx,Nojiri:2007as}. If $f(R,L_m,T)= f(R,T)$, we reobtain the field equations of the $f(R,T)$ gravity model, while the particular case $f(R,L_m,T)= f(R,L_m)$ gives the field equations of the $f(R,L_m)$ theory. Furthermore, when $f(R,L_m,T)= R$ (i.e., the Hilbert–Einstein Lagrangian), we recover the standard field equations in pure GR, namely, $R_{\mu\nu}-(1/2)g_{\mu\nu}R = 8\pi T_{\mu\nu}$.

Taking into account the covariant divergence of the field equations (\ref{eq8}), we find the non-conservation equation of the energy–momentum tensor $T_{\mu\nu}$:
\begin{align}
    \nabla^{\mu}T_{\mu\nu} =&\ \frac{1}{8\pi + f_m}\Big[ \nabla_{\nu} (L_m f_m) - T_{\mu\nu} \nabla^{\mu} f_m  \nonumber  \\
    &\left.- A_\nu - \frac{1}{2}(f_T \nabla_\nu T + f_L \nabla_\nu L_m) \right] ,  \label{conservation}
\end{align}
where we have used the fact that $\nabla^\mu R_{\mu\nu}= \nabla_\nu R/2$ and the mathematical property $(\square\nabla_\nu - \nabla_\nu\square)\phi = R_{\mu\nu}\nabla^\mu\phi$, valid for any scalar field $\phi$. Moreover, we have defined
\begin{align}
    f_m &= f_T + \frac{1}{2}f_L ,   &   A_\nu &= \nabla^\mu(f_T \tau_{\mu\nu}) .
\end{align}

The trace of the field equations leads to a second-order differential equation given by
\begin{align}
    3\square f_R+ Rf_R- 2(f- 2f_mL_m) = (8\pi+ f_m)T + f_T\tau ,
\end{align}
with $\tau$ being the trace of the tensor $\tau_{\mu\nu}$. For the particular case $f(R,L_m,T)=f(R)$, such an expression reduces to the widely known dynamical equation for the Ricci scalar in pure $f(R)$ gravity theories \cite{SotiriouF2010, FeliceT2010}. Notice that non-linear functions in $R$ lead to a non-vanishing scalar curvature in the exterior region of a compact star. For the sake of simplicity, in this work we are focused on the algebraic function originally proposed in Ref.~\cite{Haghani:2021fpx}, i.e., $f(R,L_m,T) = R + \alpha T L_m$, with $\alpha$ being a matter-geometry coupling constant. In other words, our main task will be to investigate the impact of the $\alpha T L_m$ term on the internal structure of NSs and QSs. The free parameter $\alpha$ will assume values that provide appreciable changes in the mass-radius relations.

For the above mentioned functional form, Eqs.~(\ref{eq8}) and (\ref{conservation}) reduce to
\begin{equation}
G_{\mu\nu}= \left[ 8 \pi + \frac{\alpha}{2}(T + 2 L_m) \right] T_{\mu\nu} + \alpha L_m(\tau_{\mu\nu}- L_mg_{\mu\nu}) ,
\label{eq_likeEinstein}
\end{equation}
and
\begin{widetext}
\begin{eqnarray}
\nabla^{\mu}T_{\mu\nu}= \frac{\alpha}{8 \pi +\alpha (L_m + T/2 )}\left[ \nabla_{\nu}\Big(L_m^{2} + \frac{1}{2}T L_m \Big) - T_{\mu\nu} \nabla^{\mu} \Big(L_m + \frac{T}{2} \Big) - \nabla^\mu(L_m\tau_{\mu\nu}) - \frac{1}{2}(L_m \nabla_{\nu}T + T \nabla_{\nu}L_m) \right] , \label{tov_like1}
\end{eqnarray}
\end{widetext}
respectively, where $G_{\mu\nu}$ is the usual Einstein tensor. As expected, the Einstein field equations $G_{\mu\nu}= 8\pi T_{\mu\nu}$ and conservation equation $\nabla^{\mu}T_{\mu\nu}= 0$ are recovered when $\alpha= 0$. Later we will examine the non-conservative effects of Eq.~(\ref{tov_like1}) on the internal structure of compact stars.

In the next section, we are going to obtain the system of differential equations that describe the hydrostatic equilibrium state of compact stars for the two choices of perfect fluid matter Lagrangian \cite{Bertolami2008, Faraoni2009}, i.e., $L_m= p$ and $L_m= -\rho$.

\section{Modified TOV equations}\label{sec3}

We discuss here some of the main procedures that lead to the deduction of the hydrostatic equilibrium equation in the context of $f(R,L_m,T)= R+ \alpha TL_m$ gravity model. To study compact stars, such as NSs, magnetars and other astrophysical structures, we assume these objects as being static (no rotation) and spherically symmetric stellar systems \cite{Carroll2004}. Thus, we must use the appropriate metric in a convenient coordinate system that describes the object being studied. The most general metric describing the spacetime geometry under this consideration is given by the line element
\begin{equation}
    ds^{2}=-e^{\nu(r)}dt^{2}+e^{\lambda(r)}dr^{2}+r^{2}(d\theta^{2}+\sin{\theta}^{2}d\phi^{2}),
    \label{eq10}
\end{equation}
where $\nu$ and $\lambda$ are radial functions that we want to determine based on the field equations (\ref{eq_likeEinstein}). 

In addition, we consider that the dense matter can be described as an isotropic perfect fluid represented by the following energy-momentum tensor:
\begin{eqnarray}
T_{\mu\nu}=(p + \rho)U_{\mu}U_{\nu} + pg_{\mu\nu},
\label{momentum_tensor}
\end{eqnarray}
with $p$ and $\rho$ representing the pressure and energy density of the fluid, respectively. The quantity $U_{\mu}$ is the four-velocity, which satisfies the normalization condition $U_{\mu}U^{\mu}= -1$, so it can be written as $U^\mu = e^{-\nu/2}\delta_0^\mu$. This implies that $T_\mu^\nu= \text{diag} (-\rho, p, p, p)$ and $T= -\rho+ 3p$.

As we can observe in Eq.~(\ref{EqTau}), $\tau_{\mu\nu}$ depends on the matter Lagrangian density $L_m$. Since there are two possibilities for the matter Lagrangian that lead to the energy-momentum tensor of a perfect fluid (\ref{momentum_tensor}), i.e., $L_m = p$ and $L_m = -\rho$ (see Ref.~\cite{Najera:2022lkf} and references therein for further discussion), we have a degeneracy on the form of the field equations (\ref{eq_likeEinstein}) which will have an impact on the stellar structure equations. We will adopt these two choices and analyze their implications on the main macroscopic properties of compact stars, such as mass and radius, in the context of $f(R,L_m,T)$ gravity. Let us start with the possibility $L_m = p$.

\subsection{Stellar structure equations for $L_m = p$}

Using $L_m= p$, the field equations (\ref{eq_likeEinstein}) reduce to
\begin{equation}
    G_{\mu\nu} = \left[8\pi + \frac{\alpha}{2}(5p- \rho)\right] T_{\mu\nu} - \alpha p^2g_{\mu\nu} ,
    \label{eq_field}
\end{equation}
and hence the 00 and 11 components are given respectively by
\begin{align}
    e^{-\lambda}\left(\frac{\lambda'}{r} - \frac{1}{r^2}\right) + \frac{1}{r^2} &= 8\pi \rho - \frac{\alpha}{2}(\rho - 5p)\rho + \alpha p^2,  \label{FEp00}  \\
    e^{-\lambda}\left(\frac{\nu'}{r} + \frac{1}{r^2}\right) - \frac{1}{r^2} &= 8\pi p + \frac{\alpha}{2}(3p - \rho)p,
    \label{FEp11}
\end{align}
where the prime denotes differentiation with respect to the radial coordinate $r$. In addition to the field equations, the non-conservation equation of the energy–momentum tensor (\ref{tov_like1}), for the index $\nu=1$, becomes 
\begin{equation}
    p' + \frac{\nu'}{2}(\rho + p) = -\frac{\alpha p(p'-\rho')}{16\pi + \alpha(5p - \rho)} .
    \label{eq14}
\end{equation}

Similar to the pure general relativistic scenario, we redefine the metric function $\lambda(r)$ as 
\begin{equation}
    e^{-\lambda(r)} = 1 - \frac{2 m(r)}{r},
    \label{eq15}
\end{equation}
where $m(r)$ represents the gravitational mass within a sphere of radius $r$. Rearranging Eqs.~(\ref{FEp00})-(\ref{eq14}), we get the differential equations required to describe static spherically symmetric stellar structures in the $f(R, L_m,T)= R+ \alpha TL_m$ gravity model with $L_m =p$, which are given by
\begin{align}
    \frac{dm}{dr} &= 4\pi r^{2}\rho +\frac{\alpha r^2}{2}\left[ \frac{\rho}{2}(5p - \rho) + p^2 \right] ,  \label{eq16}  \\
    \frac{dp}{dr} &= -\frac{(\rho+ p) \left[4\pi r p +\frac{m}{r^2} + \frac{\alpha r}{4}(3p-\rho)p \right]}{\Big(1 - \frac{2m}{r}\Big) \left[1 + \frac{\alpha p}{16\pi + \alpha(5p - \rho)} \Big(1-\frac{d\rho}{dp}\Big) \right]} ,  \label{eq17}
\end{align}
where we have considered a barotropic EoS in the form $p= p(\rho)$, so that $\rho' = (d\rho/dp)p'$.

\subsection{Stellar structure equations for $L_m = -\rho$}

In this case, Eq.~(\ref{eq_likeEinstein}) becomes
\begin{equation}
    G_{\mu\nu} = \left[8\pi + \frac{3\alpha}{2}( p- \rho)\right] T_{\mu\nu} - \alpha \rho^2 g_{\mu\nu} ,
\end{equation}
so that its 00 and 11 components are
\begin{align}
    e^{-\lambda}\left(\frac{\lambda'}{r} - \frac{1}{r^2}\right) + \frac{1}{r^2} &= 8\pi \rho + \frac{\alpha}{2}(3p-\rho)\rho ,  \label{FERho00}  \\
    e^{-\lambda}\left(\frac{\nu'}{r} + \frac{1}{r^2}\right) - \frac{1}{r^2} &= 8\pi p + \frac{3\alpha}{2}(p -\rho)p - \alpha \rho^2 .
    \label{FERho11}
\end{align}
On the other hand, from the non-conservation equation of the energy momentum tensor (\ref{tov_like1}), we obtain 
\begin{equation}
    p' + \frac{\nu'}{2}(\rho + p) = \frac{\alpha\left[ 4\rho\rho'+ 3p(\rho'- p') \right]}{16\pi + 3\alpha (p - \rho)} .
    \label{eq21}
\end{equation}

Similarly to the previous case, in view of Eqs.~(\ref{FERho00})-(\ref{eq21}), the modified TOV equations for the matter Lagrangian $L_m= -\rho$ take the form
\begin{align}
    \frac{dm}{dr} &= 4\pi r^2\rho + \frac{\alpha r^2}{4}(3p- \rho)\rho ,  \label{eq22}  \\
    \frac{dp}{dr} &= -\frac{(\rho+ p) \left[4\pi r p +\frac{m}{r^2} + \frac{3\alpha r}{4}(p-\rho)p - \frac{\alpha r}{2}\rho^2 \right]}{\Big(1 - \frac{2m}{r}\Big)\left\{1 + \frac{\alpha\left[3p(1-d\rho/dp) - 4\rho (d\rho/dp) \right]}{16\pi + 3\alpha(p - \rho)} \right\}} .  \label{eq23}
\end{align}

For both choices of the matter Lagrangian density, the modified TOV equations will be solved from the center at $r=0$ to the surface of the star at $r= r_{\rm sur}$ satisfying the boundary conditions:
\begin{align}
    m(0) &= 0,  &  \rho(0) &= \rho_c , 
\end{align}
where $\rho_c$ is the central energy density and $r_{\rm sur}$ is determined when the pressure vanishes. In this way, the total gravitational mass of the compact star will be given by $M\equiv m(r_{\rm sur})$. Given a specific EoS for dense matter, we will next construct the mass-radius diagrams for both sets of differential equations.

\section{Results}\label{sec4}

Here, we will discuss the impact of $f(R,L_m,T)$ theory on the internal structure of relativistic compact stars. At first we investigate the case $L_m = p$ and then the case $L_m = -\rho$, where the free parameter for both theories is $\alpha$. All our numerical results are therefore shown in terms of $\alpha$, which is given in $\mu_1= 10^{-78}\, \rm s^4/kg^2$ units (i.e., $1.46\times 10^{10}\, \rm m^2$ in geometric units) for $L_m= p$ and $\mu_2= 0.1\mu_1$ for the second choice $L_m= -\rho$. Of course, the solutions for $\alpha =0$ correspond to pure Einstein gravity.

\begin{figure*}
 \includegraphics[width=8.9cm]{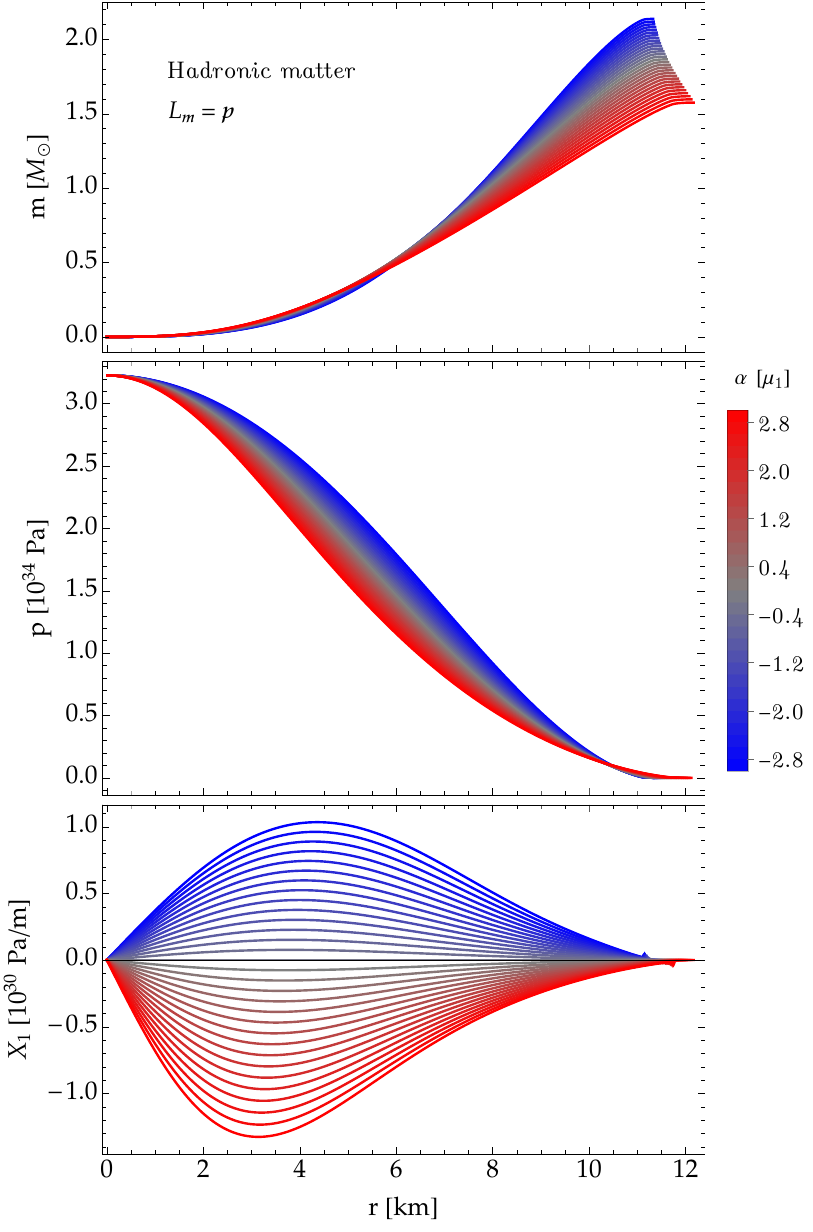}
 \includegraphics[width=8.9cm]{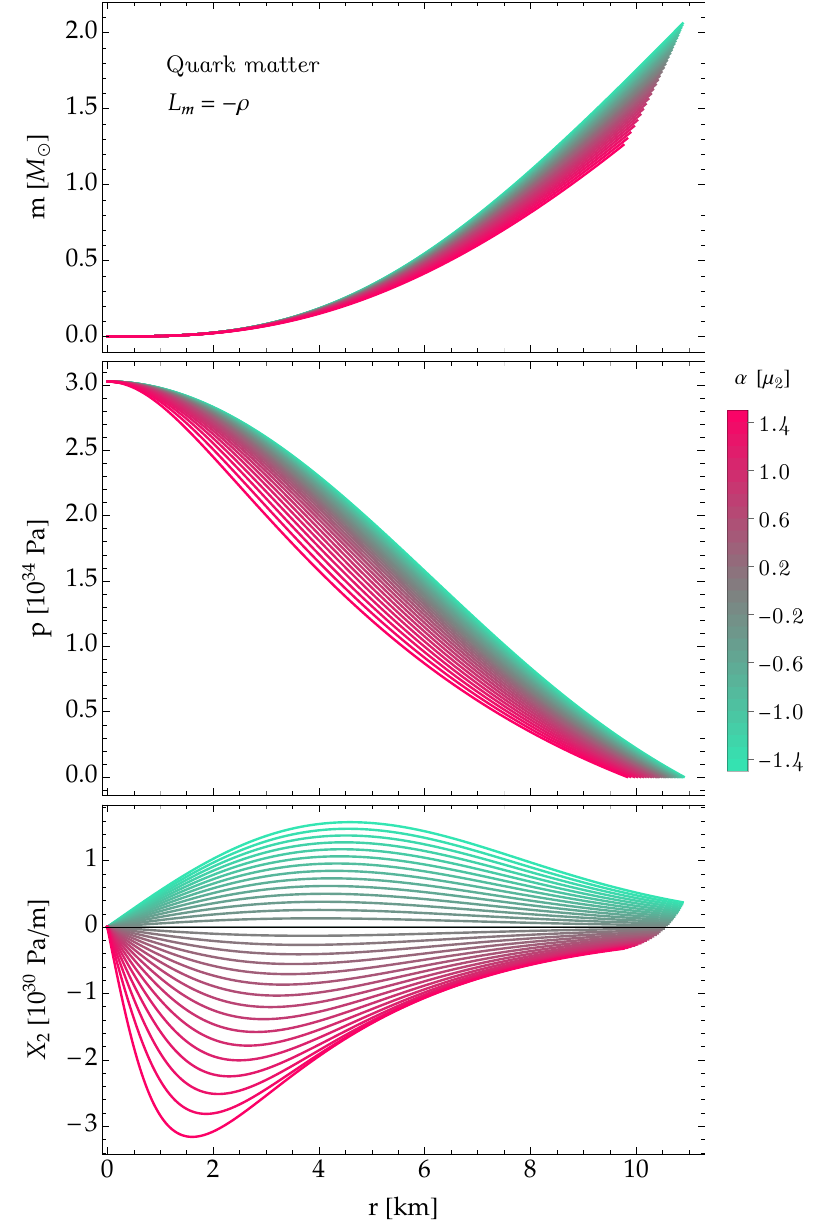}
 \caption{\label{figureRadBeha} Radial behavior of the mass distribution $m$, pressure $p$ and non-conservative term $X$ for neutron stars with $L_m= p$ (left panel) and quark stars with $L_m= -\rho$ (right panel). In both panels we have considered the same central density value $\rho_c= 1.5\times 10^{15}\, \rm g/cm^3$. Furthermore, for the free parameters we have adopted different ranges, i.e., $\alpha \in [-3.0, 3.0]\mu_1$ for $L_m= p$ and $\alpha \in [-1.5, 1.5]\mu_2$ for $L_m= -\rho$. For hadronic matter, the main consequence of the $\alpha TL_m$ term is a decrease in the total mass $m(r_{\rm sur})$ and an increase in the radius $r_{\rm sur}$ as $\alpha $ increases from its negative values. On the other hand, for quark matter, both the total mass and radius decrease with increasing $\alpha$. }  
\end{figure*}

\begin{figure*}
 \includegraphics[width=17.2cm]{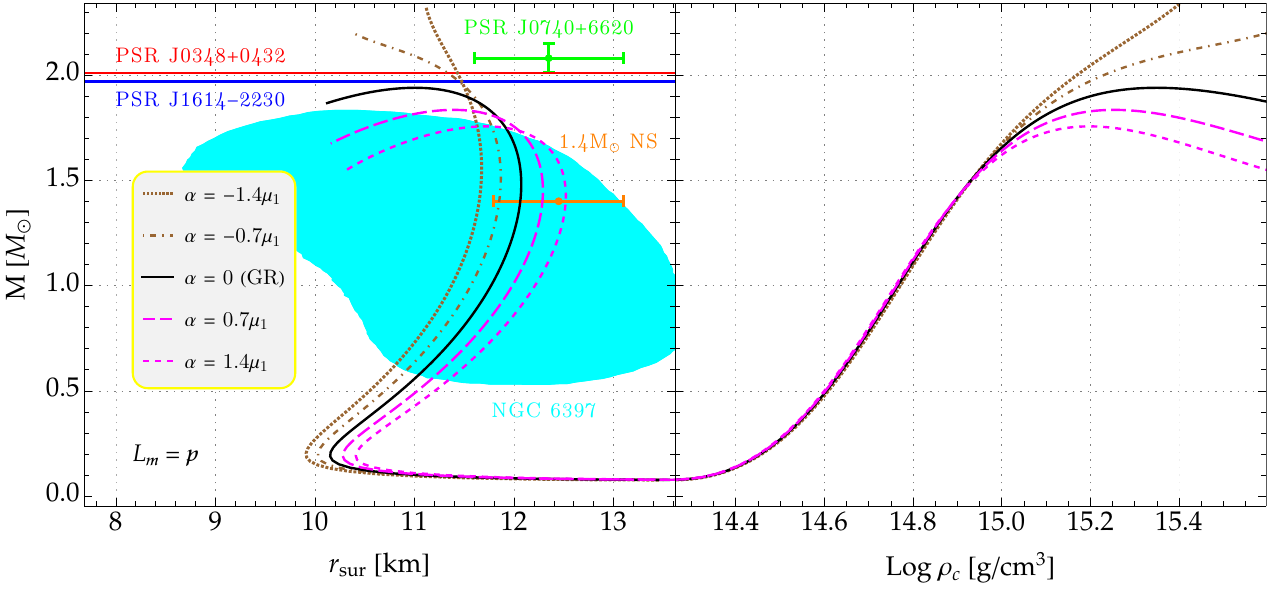}
 \includegraphics[width=17.2cm]{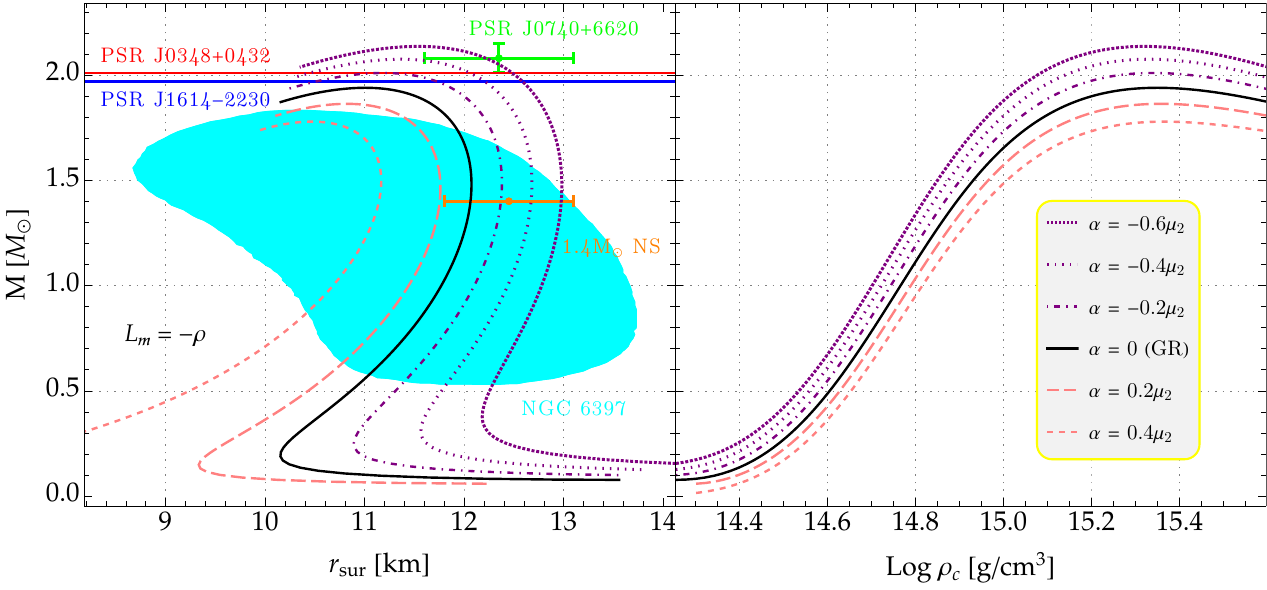}
 \caption{\label{figure1} Gravitational mass as a function of radius (left side) and central density (right side) for NSs in $f(R,L_m,T) = R + \alpha TL_m$ modified gravity using the IUFSU model, where the particular case $\alpha = 0$ corresponds to the pure GR solutions. The numerical results in the top row correspond to $L_m= p$, where $\alpha$ is given in $\mu_1= 10^{-78}\, \rm s^4/kg^2$ units (which in geometric units assumes the value $1.46\times 10^{10}\, \rm m^2$). Meanwhile, in the lower row we have chosen $L_m= -\rho$, with $\alpha$ given in $\mu_2 = 10^{-79}\, \rm s^4/kg^2$ units. The cyan region represents the NS in the quiescent low mass X-ray binary (LMXB) NGC 6397 \cite{Grindlay_2001, Guillot_2011, 10.1093/mnras/stu1449}. The blue and red lines stand for the massive NS pulsars J1614-2230 \cite{Demorest2010} and J0348+0432 \cite{Antoniadis2013}, respectively. The radius of PSR J0740+6620 (which has a gravitational mass of $2.08 \pm 0.07\, M_\odot$) from NICER and XMM-Newton Data \cite{Miller2021} is indicated by the green top dot with their respective error bars. Moreover, the orange bottom dot represents the radius estimate for a $1.4\, M_\odot$ NS \cite{Miller2021}. }  
\end{figure*}

\begin{figure*}
 \includegraphics[width=17.2cm]{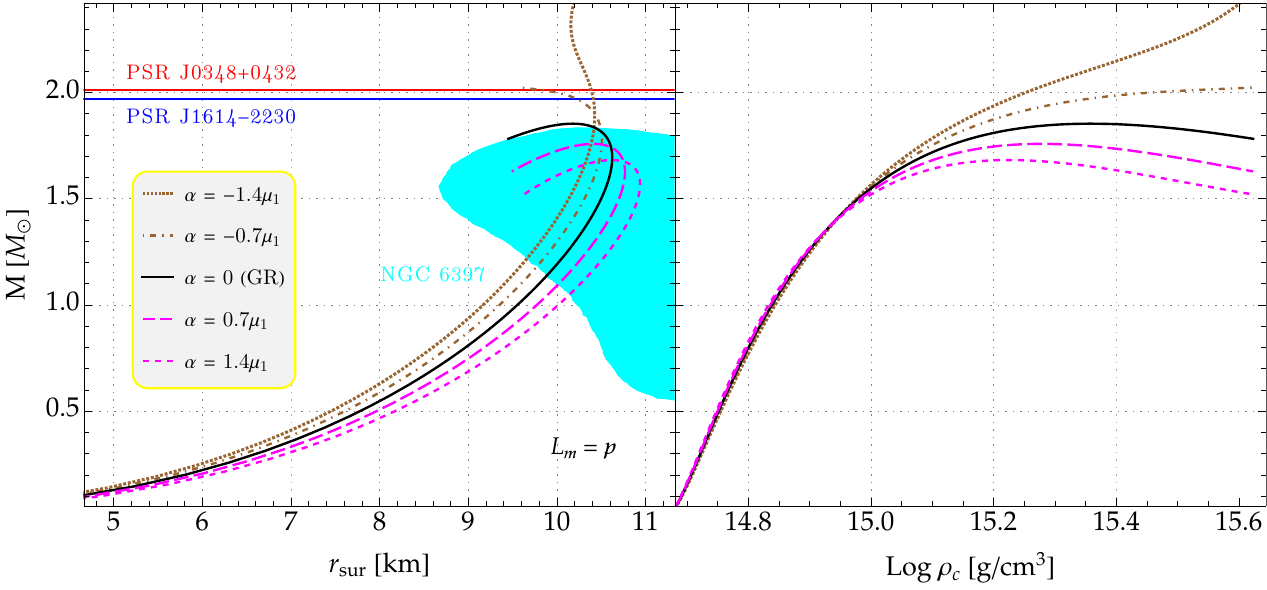}
 \includegraphics[width=17.2cm]{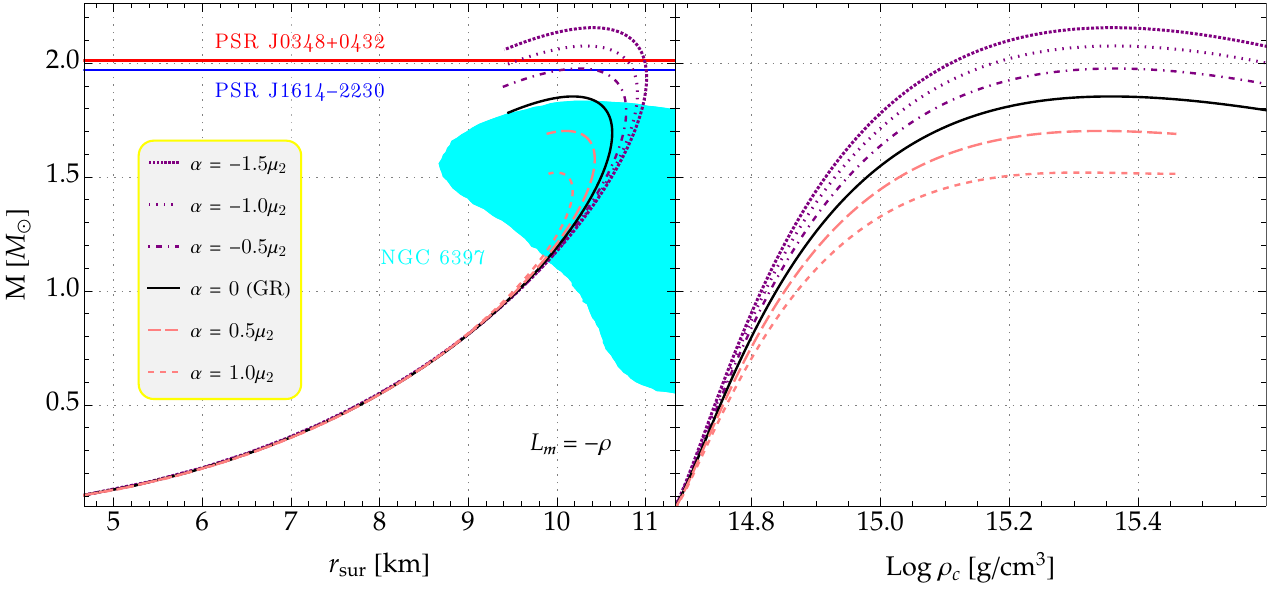}
 \caption{\label{figure2} Mass-radius diagrams (left) and mass versus central density (right) for QSs with MIT bag model EoS in $f(R,L_m,T) = R + \alpha TL_m$ gravity, where several values of $\alpha$ have been considered. As in Fig.~\ref{figure1}, the top and bottom rows correspond to $L_m= p$ and $L_m= -\rho$, respectively. }  
\end{figure*}

As input for the equations of stellar hydrostatic equilibrium, we utilized two realistic equations of state (EoS). The first one is derived from a relativistic mean-field approach, known as the IU-FSU parameterization \cite{PhysRevC.82.055803}. We opted for this parameterization due to its satisfactory capability in explaining both nuclear properties \cite{PhysRevC.99.045202} and stellar matter properties \cite{PhysRevC.93.025806}. To describe the crust of neutron stars (NSs), we adopted the complete EoS proposed by Baym, Pethick, and Sutherland (BPS) \cite{bps}. Subsequently, we compared the results obtained with the IU-FSU with those derived from an EoS for quark matter. Specifically, we employed the MIT bag model \cite{mitbag}. For a more extensive and didactic explanation on the EoS, we refer the reader to Refs.~\cite{Menezes2021, Pretel2024} and references therein.

Given a specific central density $\rho_c= 1.5\times 10^{15}\, \rm g/cm^3$, we begin our analysis by solving the stellar structure equations (\ref{eq16}) and (\ref{eq17}) for hadronic matter, where we vary the theory parameter in the range $\alpha \in [-0.3, 0.3]\mu_1$. This solution is shown in the left panel of Fig.~\ref{figureRadBeha}, where we observe that the mass profile (top plot) is strongly affected by $\alpha$ near the surface of the NS, while the pressure (middle plot) undergoes substantial changes only in the intermediate region. As a result, the mass $M$ increases (decreases) as $\alpha$ becomes more negative (positive). Note further that the radius of the star grows with increasing $\alpha$ from its negative values. However, as we will see below in the $M-\rho_c$ relations, this behavior can be reversed depending on the value of $\rho_c$.

From Eq.~(\ref{tov_like1}), it is evident that the null covariant divergence of the energy-momentum tensor is not achieved in $f(R,L_m,T)$ gravity. In order to explore the non-conservative effects of this theory on the stellar structure, we will do a graphical analysis of the right-hand term of Eq.~(\ref{tov_like1}). For the matter Lagrangian density $L_m= p$, such term becomes
\begin{equation}
    X_1 = -\frac{\alpha p(p'-\rho')}{16\pi + \alpha(5p - \rho)} ,
\end{equation}
as can be seen in Eq.~(\ref{eq14}). The lower left plot of Fig.~\ref{figureRadBeha} illustrates the magnitude of non-conservative effects along the radial coordinate of the NS. As expected, such effects become stronger as $\vert\alpha\vert$ increases. In particular, $X_1(r)$ indicates that the violation of the conservation of energy-momentum tensor has a greater repercussion in the intermediate zone, while its impact is irrelevant both at the center and at the surface of the star.

Similarly, we obtain the solution of the modified TOV equations for $L_m= -\rho$, namely, Eqs.~(\ref{eq22}) and (\ref{eq23}). Considering the same central density value, the right plot of Fig.~\ref{figureRadBeha} displays the radial profile of the mass, pressure and non-conservative term for quark matter. In this case, the non-conservative effects are quantified by
\begin{equation}
    X_2= \frac{\alpha\left[ 4\rho\rho'+ 3p(\rho'- p') \right]}{16\pi + 3\alpha (p - \rho)} ,
\end{equation}
obtained from Eq.~(\ref{eq21}). It is observed that both the mass and the radius of the quark star decrease with the increase of the parameter $\alpha$, which is varied in the interval $\alpha\in [-1.5, 1.5]\mu_2$. The non-conservative impact, measured through $X_2$, on the star has a similar behavior to $X_1$, however, it is no longer zero at the surface. This last characteristic in $X_2$ is due to the fact that the energy density is not zero at the surface of a quark star. As a consequence, given a central density, we can conclude that the internal structure of a compact star is strongly modified as the non-conservative effects are intensified by increasing $\vert\alpha\vert$.

For hadronic matter EoS in Fig.~(\ref{figure1}), we have selected the IUFSU model for both choices of the matter Lagrangian density $L_m$. We observe for the first case (upper row), that an increase for positive $\alpha$ leads to a decrease in the maximum-mass values with respect to the GR counterpart. On the other hand, for sufficiently negative values of $\alpha$ (see brown curves), it is not possible to reach the critical NS since the curve continues to grow. Concerning the radius of the star, $r_{\rm sur}$ is strongly affected by the presence of the $\alpha TL_m$ term for both positive and negative values of $\alpha$. According to the $M-\rho_c$ relation (see right panel), the gravitational mass undergoes large alterations for central densities above $10^{15}\, \rm g/cm^3$, while the changes are irrelevant below this value of $\rho_c$. In addition, the lower row of Fig.~(\ref{figure1}) presents the results for $L_m= -\rho$, where we can observe the strong impact of $\alpha$ on the mass and radius for the full range of central densities. Specifically, regardless of the value of $\rho_c$, the gravitational mass increases (decreases) for negative (positive) $\alpha$. Unlike the first choice, here it is always possible to obtain a critical configuration, that is, a NS of maximum mass on the $M-\rho_c$ curve. Another peculiar and interesting feature of this choice is that the radius does not decrease (and on the contrary, only increases as $\rho_c$ increases) when positive $\alpha$ is large enough in the small-mass region (i.e., when $M\lesssim 0.3\, M_\odot $), see for example the curve obtained for $\alpha = 0.4\mu_2$.

In Fig.~(\ref{figure2}) we exhibit our results for quark stars adopting the MIT bag model EoS. For $L_m= p$, we have a behavior similar to the NS case, that is, the mass decreases (increases) slightly for negative (positive) $\alpha$ in the low-central-density region, however, this behavior is inverted after a certain value of $\rho_c$. Indeed, it is observed that the greatest modifications take place when $\rho_c \gtrsim 10^{15}\rm g/cm^3$. On the other hand, for quark stars with $L_m= -\rho$, the effects of $\alpha$ on the $M-r_{\rm sur}$ relation are irrelevant at small masses. Nonetheless, the modifications induced by $\alpha$ become significant when $M \gtrsim 1.2\, M_\odot$.

\begin{figure*}
 \includegraphics[width=8.6cm]{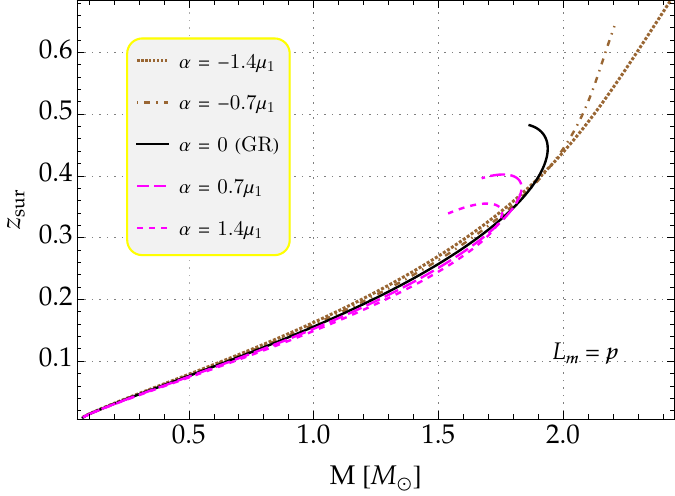}
 \includegraphics[width=8.6cm]{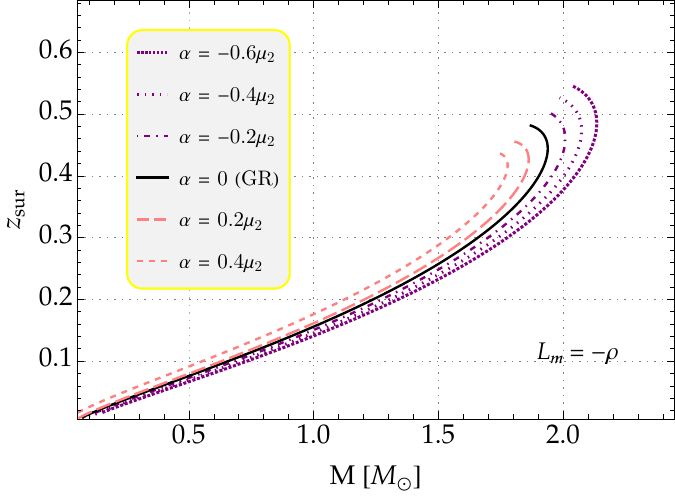}
 \caption{\label{figure2a} Surface gravitational redshift $z_{\rm sur}$ plotted against the total mass $M$ for neutron stars in $f(R,L_m,T)$ gravity with both choices of $L_m$. We have considered the same range for $\alpha$ as in Fig.~\ref{figure1}, where $\alpha = 0$ (black curve) gives the redshift of the configuration sequence in pure Einstein gravity. }  
\end{figure*}

\begin{figure*}
 \includegraphics[width=8.6cm]{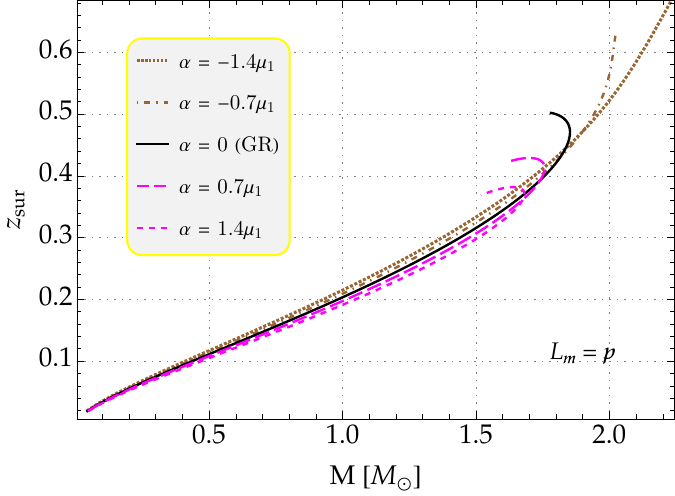}
 \includegraphics[width=8.6cm]{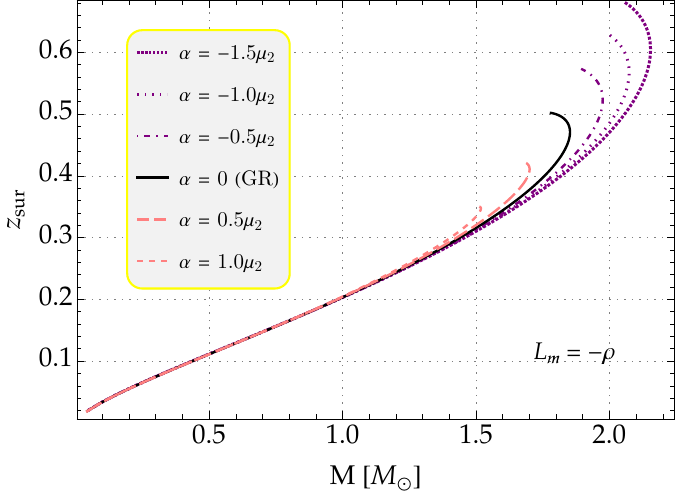}
 \caption{\label{figure2b} Gravitational redshift of light emitted at the surface of quark stars as a function of the gravitational mass for the same values of $\alpha$ as adopted in Fig.~\ref{figure2}. Remark that in the case $L_m= -\rho$, the redshift deviates substantially from the GR results only for high masses. }  
\end{figure*}

For hadronic stars, and case $L_m =p$ we can see that the results for massive stars adjust better with the constraints of the low mass -ray binary (LMXB) NGC 6397 \cite{Grindlay_2001, Guillot_2011, 10.1093/mnras/stu1449}, but the case $L_m= -\rho$ has more models outside the constrains, specifically models for positive $\alpha$. On the other this type of constraint  offer useful information for stars below approximately $2.0\, M_\odot$, in specific for stars of $1.4\, M_\odot$.  For the case of  quark stars the behavior repeats in a more strongly way. Note further that, for hadronic stars, considering the case of $L_m =p$, our findings show that for  negative values of $\alpha$ results are compatible with the massive NS pulsars J1614-2230 \cite{Demorest2010} and J0348+0432 \cite{Antoniadis2013}, contrarily to the case of positive values of  $\alpha$, which not satisfy the constraints. For QSs only the cases of negative $\alpha$ satisfy the constraints from J1614-2230 and J0348+0432.

For an isolated static spherical star, described by the line element (\ref{eq10}), the surface gravitational redshift establish a relation between mass and radius, namely $z_{\rm sur}= (1- 2M/r_{\rm sur})^{-1/2} - 1$. Here, we can also calculate the gravitational redshift of light emitted at the surface of the compact stars belonging to the several sequences shown in Figs.~\ref{figure1} and \ref{figure2}. For hadronic matter, Fig.~\ref{figure2a} exhibits the behavior of $z_{\rm sur}$ as a function of mass for for both choices of $L_m$. As expected, due to the peculiar behavior of the different curves in the $M-r_{\rm sur}$ diagrams, the redshift is strongly affected by the theory parameter $\alpha$ when $L_m= p$ for sufficiently high masses, while the changes are smaller for $L_m= -\rho$. Note further that the redshift behaves similarly for quark stars, as shown in Fig.~\ref{figure2b}.

\section{Final remarks}\label{sec5}

In summary, in this work we have investigated the phenomenology of compact stars within the framework of $f(R,L_m,T)$ gravity theories, focusing on the specific functional form $f(R,L_m,T) = R + \alpha TL_m$. The gravitational field equations for this functional form were outlined, and the non-conservation equation for the energy-momentum tensor was derived. In addition, we have considered two choices for the matter Lagrangian density, namely $L_m = p$ and $L_m = -\rho$, and examined their impact on the equilibrium structure of relativistic compact stars. The modified stellar structure equations have been derived separately for the two adoptions of $L_m$, so that the conventional TOV equations are retrieved when $\alpha= 0$. 

This study began with an introduction emphasizing the relevance of testing GR in extreme environments, such as NSs, and then introduced the $f(R,L_m,T)$ gravity model. The central idea was to examine the deviations introduced by the $\alpha TL_m$ term with respect to the GR counterpart when analyzing the most basic properties of a compact star. For two realistic EoSs of dense matter (IUFSU and MIT bag model), the resulting mass-radius diagrams were analyzed for various values of the free parameter $\alpha$. As a result, our findings indicate that the parameter $\alpha$ has a significant impact on the mass-radius relations for both NSs and QSs. We have also explored the influence of $\alpha$ on the surface gravitational redshift for the different equilibrium configuration sequences.

Specifically, for $L_m =p$, our outcomes have revealed that $\alpha$ has a prominent effect on the gravitational mass for $\rho_c \gtrsim 10^{15}\, \rm g/cm^3$, while the changes are irrelevant for central densities below this value. The maximum mass decreases as a consequence of using positive values of $\alpha$, however, it is not possible to reach a maximum-mass point (i.e., a critical configuration) when $\alpha$ is sufficiently negative. On the other hand, in the case of hadronic stars with $L_m =-\rho$, both the radius and the mass suffer significant deviations from their general relativistic value throughout the range of central densities, which does not happen for $L_m=p$. We can conclude therefore that each choice of $L_m$ has a noticeably different influence on the radius and mass of compact stars composed of hadronic matter and quark matter. Similarly to the results reported in previous studies in the strong-field regime \cite{Lobato2021, Pretel2021JCAPa, Pretel2022MPL, Bora2022, Tangphati2022PDU, Silva2023EPJC, Nashed2023, Nashed2023EPJC}, here we have shown that the non-conservative effects (product of the modification of conventional GR) play an important role within compact stars in $f(R,L_m,T)$ gravity theories.

We also confronted our results with the modern astrophysical constraint: the neutron star in the quiescent low mass X-ray binary (LMXB) NGC 6397 \cite{Grindlay_2001, Guillot_2011, 10.1093/mnras/stu1449}. We observe that all our models and chosen parameters can satisfy this important constraint. Note further that our findings for sufficiently negative values of $\alpha$ are compatible with the massive NS pulsars J1614-2230 \cite{Demorest2010} and J0348+0432 \cite{Antoniadis2013}. Remarkably, our mass-radius predictions for the choice $L_m= -\rho$ using hadronic matter favors the description of the pulsar PSR J0740+6620 from NICER and XMM-Newton Data \cite{Miller2021}, while the choice $L_m =p$ does not provide consistent results with this observational measurement. Note further that the EoSs adopted in this study do not provide maximum masses above $2M_\odot$ in pure Einstein gravity, however, $f(R,L_m,T) = R+ \alpha TL_m$ gravity with $\alpha$ sufficiently negative favors the description of different massive millisecond pulsars with masses greater than two solar masses. Thus, this work contributes to the understanding of the impact of $f(R,L_m,T)$ gravity on the internal structure of compact stars and provides valuable insights for future investigations in the field.

\section*{Acknowledgements}
JMZP acknowledges support from ``Fundação Carlos Chagas Filho de Amparo à Pesquisa do Estado do Rio de Janeiro'' -- FAPERJ, Process SEI-260003/000308/2024. COVF also make his acknowledgements to the financial suport of the productivity program  of the  Conselho Nacional de Desenvolvimento Cient\'ifico e Tecnol\'ogico (CNPq), with project number 304569/2022-4.

\end{document}